\documentclass[%
 aip,
 amsmath,amssymb,
 reprint
]{revtex4-1}

\usepackage{graphicx}
\usepackage{dcolumn}
\usepackage{bm}
\usepackage{hyperref}
\hypersetup{pdfstartview={FitH},pdfpagemode={UseNone},
            colorlinks,linkcolor=blue, citecolor=blue, urlcolor=blue,
            bookmarksopen=true, pdfnewwindow=true}
\usepackage[all]{hypcap}
\usepackage[utf8]{inputenc}
\usepackage[english]{babel}
\usepackage[T1]{fontenc}
\usepackage{mathptmx}
\usepackage{etoolbox}
\usepackage{braket}
\usepackage{siunitx}
\usepackage{amsmath}
\usepackage{amssymb}
\makeatletter
\def\@email#1#2{%
 \endgroup
 \patchcmd{\titleblock@produce}
  {\frontmatter@RRAPformat}
  {\frontmatter@RRAPformat{\produce@RRAP{*#1\href{mailto:#2}{#2}}}\frontmatter@RRAPformat}
  {}{}
}%
\makeatother
\draft 
\begin{document}

\preprint{}

\title[Generation of Tunable Entanglement from Thin-Film Lithium Niobate]{Generation of Tunable Entanglement from Thin-Film Lithium Niobate}

\author{Saniya Shinde}
\affiliation{Institute of Applied Physics, Abbe Center of Photonics, Friedrich Schiller University Jena, Albert-Einstein-Straße 15, 07745 Jena, Germany}
\affiliation{ARC Center of Excellence for Transformative Meta-Optical Systems (TMOS), Department of Electronic Materials Engineering, Research School of Physics, Australian National University, Canberra, ACT 2601, Australia}
\email{saniya.shinde@uni-jena.de}

\author{Maximilian A. Weissflog}
\affiliation{Institute of Applied Physics, Abbe Center of Photonics, Friedrich Schiller University Jena, Albert-Einstein-Straße 15, 07745 Jena, Germany}

\author{Shaun Lung}
\affiliation{Institute of Applied Physics, Abbe Center of Photonics, Friedrich Schiller University Jena, Albert-Einstein-Straße 15, 07745 Jena, Germany}

\author{Elkin A. Santos}
\affiliation{Institute of Applied Physics, Abbe Center of Photonics, Friedrich Schiller University Jena, Albert-Einstein-Straße 15, 07745 Jena, Germany}

\author{Jinyong Ma}
\affiliation{ARC Center of Excellence for Transformative Meta-Optical Systems (TMOS), Department of Electronic Materials Engineering, Research School of Physics, Australian National University, Canberra, ACT 2601, Australia}
\affiliation{Institute of Quantum Precision Measurement, State Key Laboratory of Radio Frequency Heterogeneous Integration, College of Physics and Optoelectronic Engineering, Shenzhen University, Shenzhen 518060, P. R. China}

\author{Tongmiao Fan}
\affiliation{ARC Center of Excellence for Transformative Meta-Optical Systems (TMOS), Department of Electronic Materials Engineering, Research School of Physics, Australian National University, Canberra, ACT 2601, Australia}

\author{Anna Fedotova}
\affiliation{Institute of Applied Physics, Abbe Center of Photonics, Friedrich Schiller University Jena, Albert-Einstein-Straße 15, 07745 Jena, Germany}
\affiliation{Institute of Solid State Physics, Friedrich Schiller University Jena, Helmholtzweg 3, 07743 Jena, Germany}

\author{Sina Saravi}
\affiliation{Heinz Nixdorf Institute, Paderborn University, Fürstenallee 11, 33102 Paderborn, Germany}
\affiliation{Department of Electrical Engineering and Information Technology, Paderborn University, Warburger Straße 100, 33098 Paderborn, Germany}
\affiliation{Institute for Photonic Quantum Systems (PhoQS), Paderborn University, Warburger Str. 100, 33098 Paderborn, Germany}
\affiliation{Institute of Applied Physics, Abbe Center of Photonics, Friedrich Schiller University Jena, Albert-Einstein-Straße 15, 07745 Jena, Germany}

\author{Andrey A. Sukhorukov}
\affiliation{ARC Center of Excellence for Transformative Meta-Optical Systems (TMOS), Department of Electronic Materials Engineering, Research School of Physics, Australian National University, Canberra, ACT 2601, Australia}

\author{Frank Setzpfandt}
\affiliation{Institute of Applied Physics, Abbe Center of Photonics, Friedrich Schiller University Jena, Albert-Einstein-Straße 15, 07745 Jena, Germany}
\affiliation{Fraunhofer Institute for Applied Optics and Precision Engineering IOF, Albert-Einstein-Straße 7, 07745 Jena, Germany}

%%%%%%%%%%%%%%%%%%%%%%%%%%%%%%%%%%%%%%%%%%%%%%%%%%%%%%%%%%%%%%%%%%%%%%%%%%%%%%%%%

\begin{abstract}
Entangled photon pairs play a major role in various modern technologies such as quantum imaging, communication, and computing. Conventional photon-pair sources are often based on spontaneous parametric down-conversion in bulk nonlinear crystals. 
Recent advances have also shown entangled photon-pairs from transition metal dichalcogenide thin-films, however, these materials are not widely available and are not compatible with existing fabrication capabilities. We present a new thin-film lithium niobate source of polarization-entangled photon pairs at the telecom wavelength that requires no additional optical elements for entanglement generation and allows for easy application using the existing lithium niobate fabrication technologies. 
We demonstrate tunable entanglement generation using the three-fold rotational crystal symmetry of lithium niobate, allowing the generation of different maximally entangled Bell states or completely separable states depending on the polarization of the pump beam. 
  
\end{abstract}

\maketitle

\section{\label{sec:level1}Introduction}

\noindent Entangled photon pairs are an essential part of many applications of quantum technologies, e.g., in quantum imaging \cite{gilaberte_basset_perspectives_2019}, quantum computing \cite{obrien_optical_2007}, and quantum communication\cite{gisin_quantum_2007}. The generation of photon pairs is often based on spontaneous parametric down-conversion (SPDC) in bulk nonlinear crystals with a second-order nonlinearity, such as $\beta$-barium-borate (BBO) \cite{kwiat_new_1995}, lithium niobate (LN)\cite{anwar_entangled_2021}, or periodically poled potassium titanyl phosphate (ppKTP) \cite{kim_phase-stable_2006}. However, generating entangled pairs requires at least two different superimposed pathways for SPDC, i.e., the simultaneous possibility to generate pairs of distinct signal and idler photons entangled in one degree of freedom. In many common generation schemes, these pathways are isolated from a more complex state using filtering \cite{kwiat_new_1995} or they are realized separately by two distinct SPDC interactions, which are subsequently interfered. Examples for the latter approach include entanglement generation in a Sagnac loop \cite{shi_generation_2004}, a linear double-pass configuration\cite{steinlechner_phase-stable_2013}, or a crossed-crystal arrangement \cite{kwiat_ultrabright_1999} where the nonlinear axes of the two used nonlinear crystals are perpendicular, thus each crystal generates photon pairs of different polarizations. Each of these methods requires additional optical elements besides the nonlinear crystals to enable entangled state generation with high fidelity.

In recent years, photon-pair sources based on thin-film nonlinear media have emerged \cite{okoth_microscale_2019, santiago-cruzEntangledPhotonsSubwavelength2021, guo_ultrathin_2023, sultanov_temporally_2024}. Here, phase-matching constraints are lifted, which allows one to use material properties that cannot readily be exploited in bulk nonlinear media. This has enabled the generation of polarization-entangled photon pairs in a single propagating mode without any additional optical elements \cite{weissflog_tunable_2024,feng_polarization-entangled_2024, lu_counter-propagating_2025,kallioniemiVanWaalsEngineering2025,sultanov_flat-optics_2022,guoPolarizationEntanglementEnabled2024,liang_tunable_2025,santos_entangled_2024}. 

\begin{figure*}
  \includegraphics[width=\textwidth]{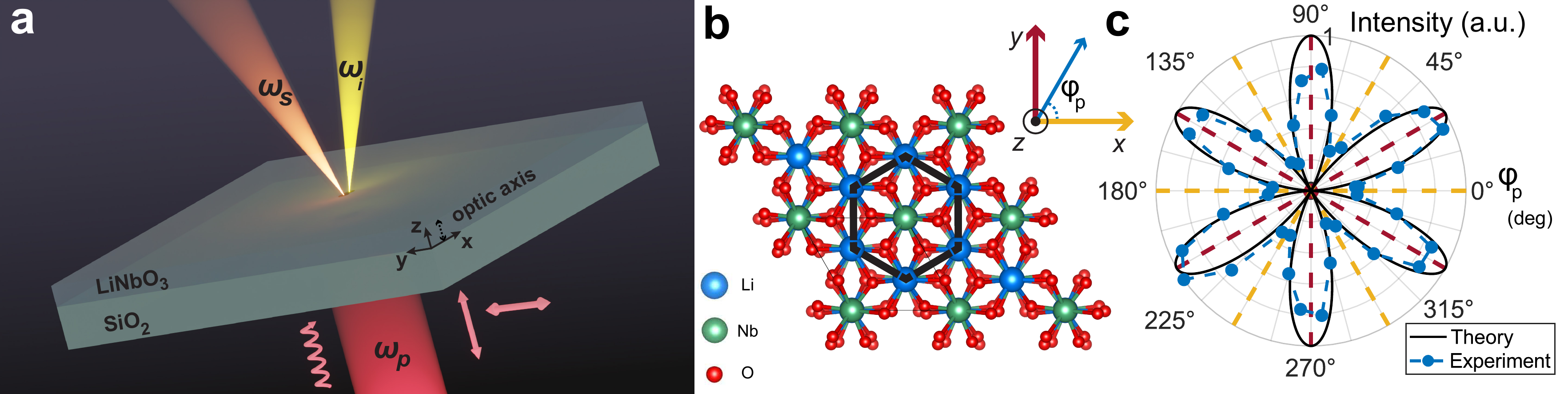}
  \caption{(a) Schematic of entanglement generation in z-cut LN. Sample is pumped from the substrate side with various pump polarizations. (b) Top-view of the crystal structure of z-cut lithium niobate (chemical formula LiNbO$_3$, rendered using VESTA3\cite{mommaVESTA3Threedimensional2011}). The black hexagon is marked to highlight the the six-fold rotational symmetry of the crystal structure. The pump polarization angle $\varphi_\mathrm{p}$ is shown with respect to crystalline \textit{x}- and \textit{y}-axes. (c) Polarization resolved. normalized second-harmonic generation intensity pattern with co-rotating half-wave plate and analyzer. Yellow and red dashed lines indicate \textit{x}- and \textit{y}- directions in the crystal. The theoretical dependence of $ I_{SHG} \propto \mathrm{sin}^2(3\varphi_\mathrm{p})$ is shown by the solid black curve.}
  \label{fgr:1}
\end{figure*}

A particularly simple approach, based on a hitherto unused material symmetry, first theoretically proposed in Ref.~\onlinecite{visser_polarization_2002}, was followed in Ref.~\onlinecite{weissflog_tunable_2024}. Here, the transition metal dichalcogenide (TMD) 3R-MoS$_2$, exhibiting a $C_{3v}$ symmetry was exploited. The second-order nonlinear tensor associated with this symmetry class facilitates the direct generation of different, maximally polarization-entangled photon-pair states, which can be tuned by a simple rotation of the linear pump polarization. For this material class, it was furthermore demonstrated that the degree of entanglement can be controlled from maximally entangled to completely factorizable states by changing the pump polarization from linear to circular \cite{feng_polarization-entangled_2024}. These materials show a very high nonlinearity, however, they are constrained by their transparency region. Broadband applications, in particular, would be challenging with TMDs due to photoluminescence and excitonic resonances in the visible and near-infrared (NIR) wavelengths. Furthermore, although progress has been made in the growth of TMDs, obtaining large area, monocrystalline TMD films with uniform thickness is still challenging. Therefore, for fabrication of high-quality, monocrystalline TMD sheets, manual exfoliation is still used, which is not an inherently scalable method even in combination with quasi-phase matching.

Interestingly, the well established nonlinear material LN exhibits the same $C_{3v}$ crystal symmetry as the aforementioned transition metal dichalcogenides. LN is a very attractive material for integrated and quantum photonics due to a number of advantageous properties, like a broad transparency window (350-4500 nm), a comparably large nonlinearity, and strong electro-optic response \cite{fedotova_lithium_2022}. Thin-film lithium niobate is commercially available with uniform thickness, also in wafer-scale sizes, and has standard lithographic fabrication methods.
It is regularly used as a source of photon pairs: as periodically poled bulk LN (PPLN) \cite{anwar_entangled_2021}, as a waveguide \cite{zhu_integrated_2021}, as an unstructured thin film \cite{okoth_microscale_2019}, and in the form of nanostructured metasurfaces \cite{santiago-cruz_photon_2021,ma_polarization_2023, weissflog_directionally_2024}. Generation of polarization-entangled states was demonstrated in bulk LN crystals \cite{anwar_entangled_2021}, LN waveguides \cite{kim_integrated_2025}, and proposed in LN metasurfaces \cite{ma_polarization_2023, mo_polarization-entangled_2025, shi_enhanced_2025}. All of these sources are using the largest elements $d_{33}$ and $d_{31}$ of LN's nonlinear tensor and are therefore not able to exploit the $C_{3v}$ symmetry. Instead, they need sophisticated nanostructuring or additional optical elements to generate entanglement.

In this work, we show that it is possible to directly generate tunable polarization entanglement in photon pairs from unstructured thin-film lithium niobate using its $d_{22}$ tensor component. As shown in Fig.~\ref{fgr:1}a, we use a z-cut LN film, where the optic axis of the uniaxial LN crystal is oriented perpendicular to the surface and parallel to the used pump propagation direction. The hexagonal structure of the crystalline lattice as seen along the pump propagation direction is shown in Fig.~\ref{fgr:1}b. In this way, we can exploit the $C_{3v}$ symmetry of the crystal to generate entanglement without the need for any additional patterning or post-processing.
\begin{figure*}
  \includegraphics[width=0.8\textwidth]{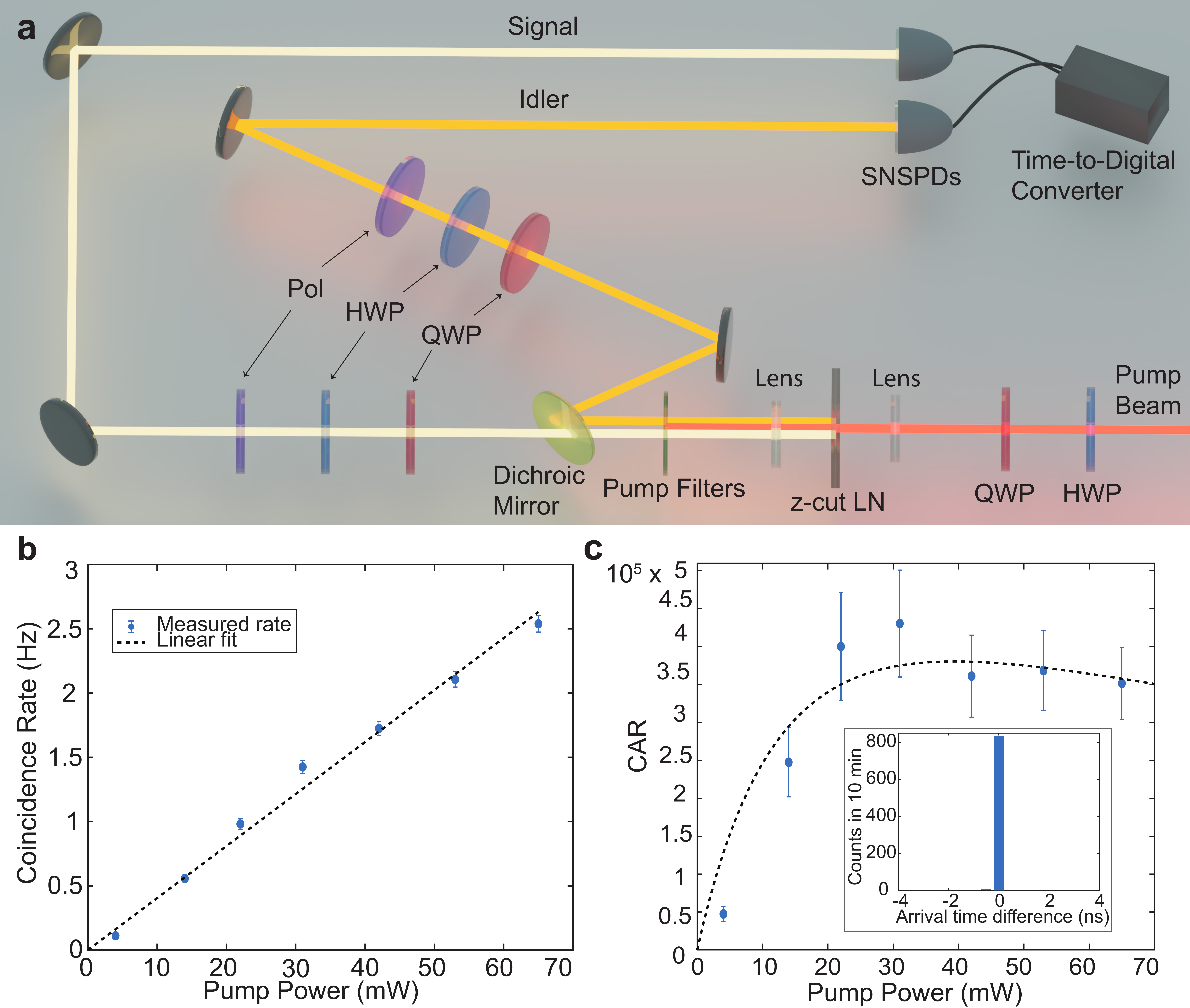}
  \caption{(a) Experimental setup for quantum state tomography. Photon pairs are separated by a dichroic mirror into two beam paths. Each path contains a quarter-wave plate (QWP), a half-wave plate (HWP), and a fixed polarizer aligned along the vertical direction (Pol) and detected using single-mode fibers connected to superconducting nanowire single photon detectors (SNSPDs). (b) SPDC rate as a function of pump power. Error bars are calculated using Poissonian statistics. Dashed black line shows the theoretical fit. (c) Coincidence-to-accidental ratio (CAR) as a function of power. Error bars are calculated using Poissonian statistics. Dashed black line shows theoretical fit of CAR using rate of single counts. Inset: coincidence histogram for measurement with pump power $P=\SI{31}{mW}$, bin width \qty{500}{\pico\second} integrated over 10 min.}
  \label{fgr:2}
\end{figure*}

\section{Results and Discussion}

\noindent The nonlinear tensor of crystals belonging to the $C_{3v}$ symmetry class has a number of non-zero second-order nonlinear tensor components \cite{boyd_nonlinear_2020}:
$\chi_{xzx}^{(2)} = \chi_{yzy}^{(2)}, \chi_{xxz}^{(2)} = \chi_{yyz}^{(2)}, \chi_{zxx}^{(2)} = \chi_{zyy}^{(2)}, \chi_{zzz}^{(2)}, \chi_{yyy}^{(2)} = -\chi_{yxx}^{(2)} = -\chi_{xxy}^{(2)} = -\chi_{xyx}^{(2)}$.

Here, we examine a configuration where the pump, signal and idler beams propagate along the crystalline \textit{z}-direction and can thus access the in-plane $d_{22}$ tensor components $\chi^{(2)}_{yyy}=-\chi^{(2)}_{yxx}=-\chi^{(2)}_{xxy}=-\chi^{(2)}_{xyx}$. These have a magnitude of \qty{4.2}{\pico\meter/\volt} \cite{roberts_simplified_1992}, smaller than the tensor elements typically used for nonlinear three-wave mixing in LN. 
However, the symmetry of z-cut LN and the accessible tensor elements allow us to directly generate and switch between maximally entangled Bell states by choosing a suitable linear pump polarization. Additionally, it offers control over the degree of entanglement by changing the ellipticity of the pump polarization. 

For further explanations of the generation mechanism, we label the polarization direction along the \textit{y}-axis of the crystal as V-polarization and along the crystalline \textit{x}-axis as H-polarization. When the pump photon is linearly polarized along the \textit{y}-axis, we are exciting the tensor components $\chi^{(2)}_{yyy}$ and $\chi^{(2)}_{xxy}$. Since these are equal in magnitude, the generated photon pairs can be either \textit{xx}- or \textit{yy}-polarized with equal probability. Assuming that signal and idler can be distinguished in their wavelength and are indistinguishable in their spatial properties, the two generation possibilities interfere and generate the entangled output state $\ket{\Phi^-} = \frac{1}{\sqrt{2}} \left(\ket{HH} - \ket{VV}\right)$.
Similarly, when the pump photon is polarized along the \textit{x}-axis, we are exciting the tensor components $\chi^{(2)}_{xyx}$ and $\chi^{(2)}_{yxx}$, which are equal in magnitude and sign, leading to the observation of the Bell state $\ket{\Psi^+} = \frac{1}{\sqrt{2}} \left(\ket{HV} + \ket{VH}\right)$. Assuming a plane-wave excitation, all other nonlinear conversion processes involving \textit{z}-polarized modes are suppressed.

In our experiments, we use a LN thin film of \qty{8}{\micro\meter} thickness. 
This corresponds to the coherence length for SPDC with a pump around $\sim$ \qty{788}{\nano\meter} and signal/idler photon pairs in the telecom wavelength range around $\sim$ \qty{1576}{\nano\meter} (see Supplementary Section~1 for more details). With this we maximize the generation rate before the onset of detrimental effects related to phase mismatches.

\begin{figure*}
  \includegraphics[width=1\textwidth]{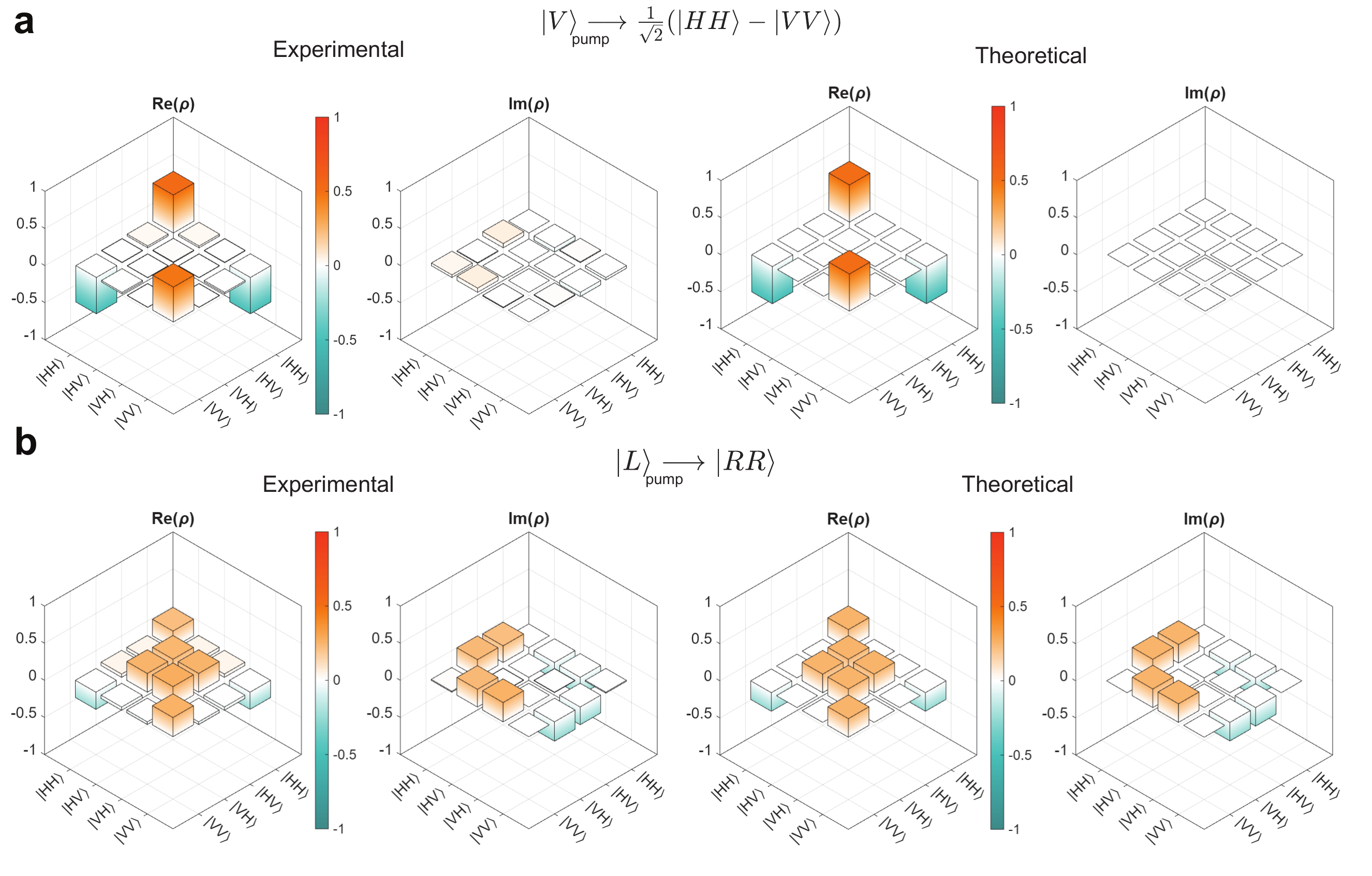}
  \caption{(a) Experimental and theoretical density matrices for V-polarized pump. For the experimental results, concurrence is $\mathrm{C} = 0.97 \pm 0.013$ with state fidelity $\mathrm{F(\Phi^-)} = 0.98$. (b) Experimental and theoretical density matrices for LCP pump. For the experimental results, concurrence is $\mathrm{C} = 0.16 \pm 0.037$ with state fidelity $\mathrm{F(RR)} = 0.97$. Experimental density matrices were obtained using quantum state tomography and maximum likelihood estimation.}
  \label{fgr:3}
\end{figure*}

\subsubsection{Second-Harmonic Measurements} 
\noindent In order to identify the crystalline \textit{x}- and \textit{y}-axes in our sample, we perform second-harmonic measurements for a fundamental harmonic wavelength of $\lambda_{FH}$ = \qty{1576}{\nano\meter} and second-harmonic wavelength $\lambda_{SH}$ = \qty{788}{\nano\meter}. 
We place a half-wave plate before the sample to control the polarization angle $\varphi_\mathrm{p}$ of the input fundamental harmonic and a polarizer after the sample, which acts as the analyzer for the second harmonic. A schematic of the experimental setup is shown in the Supplementary Fig.~S2. Fig.~\ref{fgr:1}c shows the measured second harmonic generation (SHG) intensity dependence on $\varphi_\mathrm{p}$ where the SH analyzer is always aligned parallel to the fundamental beam polarization direction. The SHG intensity $I_{SHG}$ shows a six-fold pattern $I_{SHG} \propto \mathrm{sin}^2(3\varphi_\mathrm{p})$, characteristic for a nonlinear crystal with $C_{3v}$ symmetry. The maximum is oriented along the \textit{y}-axis of the crystal as shown with the red dashed line in Fig.~\ref{fgr:1}c. The yellow dashed lines in Fig.~\ref{fgr:1}c highlight the \textit{x}-axis direction of z-cut LN.

\subsubsection{SPDC Characterization Measurements}
\noindent Subsequently, we perform SPDC measurements determining the photon-pair rate and coincidence-to-accidental ratio (CAR) using a Hanbury Brown-Twiss setup. %sketched in Fig.~\ref{fgr:2}a. 
Here, the two photons of a pair are collected using only the transmission arm of the setup shown in Fig.~\ref{fgr:2}a and are then separated by a probabilistic fiber beamsplitter. Subsequently they are directed to two single-photon detectors connected to a time-correlator. With this configuration, the pair-emission rate can be measured through the temporal correlation of the detectors. A schematic of this experimental configuration is shown in the Supplementary Fig.~S2. We observe that the SPDC rate is directly proportional to the pump power (Fig.~\ref{fgr:2}b), a hallmark of photon-pair generation. The coincidence-to-accidental ratio (CAR) strongly exceeds the value of two associated with classical thermal light, further evidencing the generation of photon pairs. The CAR shows the typical inverse proportionality $\mathrm{CAR}\propto P^{-1}$ with respect to the pump power $P$ for values $P\gtrsim\SI{20}{mW}$, see Fig.~\ref{fgr:2}c. For lower powers, the single count rate is dominated by the environmental dark counts rather than detected photons (see Supplementary Section~2 for a more detailed discussion), resulting in a deviation from the $\mathrm{CAR}\propto P^{-1}$ dependence. One of the SPDC histograms obtained for pump power \qty{31}{\milli\watt} is shown as inset of Fig.~\ref{fgr:2}c. For the pump power of \qty{65}{\milli\watt} we observe an SPDC rate of 2.54\;$\pm$ 0.06\;Hz and CAR of $\mathrm{CAR}= (1.17 \pm 0.16) \times 10^5$. These measurements were carried out for a coincidence window width of $\tau=\SI{500}{ps}$ and integration time $\SI{10}{min}$.

\begin{figure*}
  \includegraphics[width=0.7\textwidth]{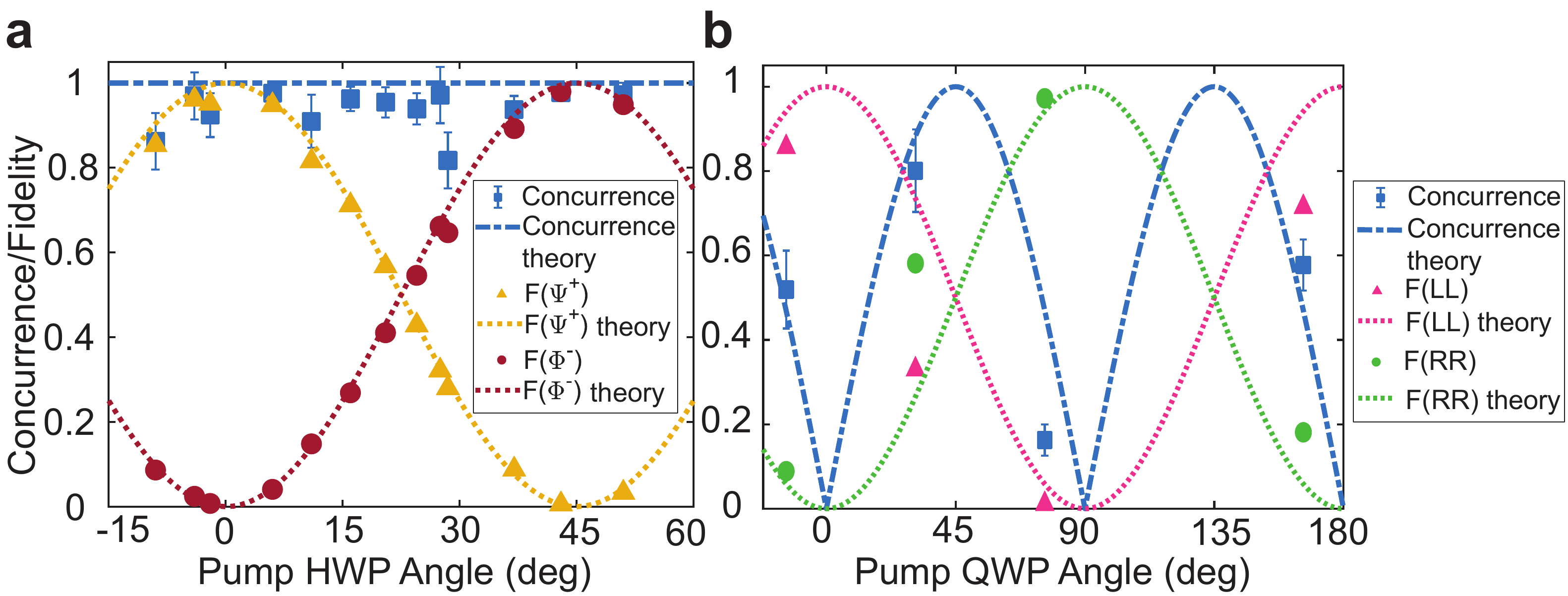}
  \caption{(a) Concurrence and Bell state fidelities as a function of pump half-wave plate angle. Dashed blue line shows the theoretically predicted concurrence and dashed red and yellow lines indicate the theoretical Bell state fidelities. (b) Concurrence and non-entangled state fidelities as a function of pump quarter-wave angle, for a fixed pump half-wave plate angle corresponding to the pump polarization angle $\varphi_\mathrm{p}=\SI{45}{\degree}$.Dashed blue line shows the theoretically predicted concurrence and dashed pink and green lines indicate the theoretical non-entangled state fidelities.}
  \label{fgr:4}
\end{figure*}

\subsubsection{Quantum State Tomography}
\noindent For determining the polarization quantum state, we use the experimental configuration shown in Fig.~\ref{fgr:2}a. Here, the photons are deterministically split by a dichroic mirror into two frequency modes, which allow us to distinguish signal and idler photons. Subsequently, both photons are sent through a variable, elliptical polarizer consisting of a quarter- and half-wave plate and a linear polarizer. By consecutive projections into different basis combinations, the polarization density matrix can be tomographically reconstructed \cite{jamesMeasurementQubits2001}.
 
As a measure of the degree of entanglement, we calculate the concurrence $\mathrm{C}$, where $\mathrm{C}=1$ corresponds to an entangled and $\mathrm{C}=0$ to a separable state, respectively.
For pump polarization along \textit{x} (H-pol), the output photon-pair state has a measured concurrence of $\mathrm{C} = 0.92 \pm 0.05$ and fidelity $\mathrm{F} = 0.95$ with respect to the theoretically expected $\ket{\Psi^+}$ entangled state. For pump polarization along \textit{y} (V-pol), the output photon-pair state has concurrence $\mathrm{C} = 0.97 \pm 0.01$ and fidelity $\mathrm{F} = 0.98$ with respect to the $\ket{\Phi^-}$ entangled state (experimentally retrieved and theoretically expected density matrix shown in Fig.~\ref{fgr:3}a). 

In addition to these maximally-entangled states, z-cut LN can also produce completely non-entangled states, simply by changing the pump polarization. If, instead of a linearly polarized input, we use a circularly polarized pump, we obtain a separable state with the opposite spin angular momentum in accordance with conservation of spin angular momentum\cite{simon_second-harmonic_1968}. 
A left-circularly polarized (LCP) pump produces two right-circularly polarized (RCP) photons ($\ket{RR}$ quantum state) with fidelity $\mathrm{F} = 0.97$. The corresponding measured density matrix is shown in Fig.~\ref{fgr:3}b in comparison with the theoretical expectation in this case. With an RCP pump we obtain the $\ket{LL}$ quantum state with fidelity $\mathrm{F} = 0.86$ (not shown).

In the final step, we demonstrate the flexibility in tuning the polarization quantum state, in particular that the degree of entanglement can be continuously tuned between separable and fully entangled states. For this we change the polarization ellipticity of the pump beam. Formally, the pump polarization can be expressed in the orthogonal \textit{x}- and \textit{y}-basis states, which enables writing any arbitrary elliptically-polarized pump photon as: $\psi_\mathrm{p} = \mathrm{cos}(\varphi_\mathrm{p})\ket{x} + e^{i \delta} \mathrm{sin}(\varphi_\mathrm{p}) \ket{y}$. Here, $\varphi_\mathrm{p}$ is the linear pump polarization angle and $\delta$ is the relative phase difference between the $x$- and $y$-polarization components. The general photon-pair state as a function of the pump polarization parameters $\varphi_\mathrm{p}$ and $\delta$ can be obtained by superimposing the $\ket{\Psi^+}$ and $\ket{\Phi^-}$ states that are generated for linear $x$- or $y$-polarized excitation. As a result we find
\small
\begin{equation}
    \ket{\psi (\varphi_\mathrm{p}, \delta_\mathrm{p})} = \frac{\mathrm{cos}(\varphi_\mathrm{p})}{\sqrt{2}} (\ket{HV} + \ket{VH}) + e^{i \delta} \frac{\mathrm{sin}(\varphi_\mathrm{p})}{\sqrt{2}} (\ket{HH} - \ket{VV}). 
    \label{eq:1}
\end{equation}
\\
\normalsize

A more detailed explanation of Eq.~\eqref{eq:1} is provided in Supplementary Section~3. We first demonstrate experimentally that the polarization state $\ket{\psi(\varphi_\mathrm{p}, 0)}$ remains maximally entangled for any linear pump polarization. For this, we vary the angle of the half-wave plate in the pump beam path to change the linear pump angle $\varphi_\mathrm{p}$. The experimentally obtained concurrence and fidelity of the output photon-pair states with respect to the $\ket{\Phi^-}$- and $\ket{\Psi^+}$-states is shown in Fig.~\ref{fgr:4}a. The concurrence stays at high values close to one, independent of the polarization orientation, while the fidelity with respect to the Bell states varies sinusoidally. This matches the theoretically predicted dependencies based on  Eq.~\eqref{eq:1}, which are shown as dashed lines in Fig.~\ref{fgr:4}a. Overall, we find very good agreement between the experimental results and the theoretical model. 

We then add a quarter-wave plate to the pump beam path to control the phase delay $\delta_p$ between the $x$- and $y$-polarization components of the pump beam while keeping the pump half-wave plate angle fixed at \qty{22.5}{\degree}. This corresponds to a fixed angle $\varphi_\mathrm{p}=\SI{45}{\degree}$. We vary the pump quarter-wave plate angle and measure the concurrence of the states $\ket{\psi(\SI{45}{\degree}, \delta_\mathrm{p})}$ obtained and their fidelities with respect to the $\ket{LL}$ and $\ket{RR}$ quantum states, see Fig.~\ref{fgr:4}b. We find that the concurrence can indeed be tuned continuously in the range $0\lesssim C\lesssim1$, purely based on the pump ellipticity, and follows the theoretically predicted trend as shown by the dashed blue line in Fig.~\ref{fgr:4}b. The dashed green and pink curves in Fig.~\ref{fgr:4}b indicate the theoretically predicted fidelity with respect to the two separable states $\ket{RR}$ and $\ket{LL}$, respectively, obtained from  Eq.~\eqref{eq:1}.

\section{Conclusion}

\noindent With our results, we have shown that leaving the track of phase-matched nonlinear bulk optics allows the use of well-established nonlinear materials such as lithium niobate (LN) for photon-pair generation in an unconventional way. In particular, we have shown the generation of maximally entangled Bell states directly from thin-film lithium niobate without any additional optical components. This is enabled by the three-fold rotational symmetry of LN's crystalline structure, which gives rise to a nonlinear tensor that intrinsically allows generation of entangled pairs, but could not be exploited in earlier works that focused on using the largest elements of the nonlinear tensor. In addition, we have demonstrated tunability of polarization entanglement depending on the pump polarization. For linearly polarized pump states, we obtained high-fidelity Bell states $\ket{\Phi^-}$ and $\ket{\Psi^+}$ for V- and H-polarized pump, respectively. For a circularly polarized pump, we observed a complete loss of entanglement, with an RCP or LCP pump producing the separable state $\ket{LL}$ or $\ket{RR}$, respectively. Furthermore, we showed that for an elliptically polarized pump the state concurrence can be continuously tuned.

While 2D material entangled photon-pair sources with similar C$_{3v}$ symmetry are shown to have higher nonlinearity (see Supplementary Section~4 for a comparison), z-cut LN offers a unique advantage for applications with it's low background counts, large optical transparency bandwidth and scalability. In the future, the efficiency of pair-generation can be increased by using this source in combination with a cavity \cite{slattery_background_2019,kongDoublyResonantSecondharmonic2023} or using resonant nanostructures where modal interference can be used to tailor the directionality of the photon pairs produced \cite{weissflogNonlinearNanoresonatorsBell2024}. 
Another possible method to increase the pair-generation rate is to use stacks of periodically rotated z-cut LN crystal sheets, which introduces quasi-phase matching in the $z$-direction, such that the thickness of the crystal is no longer limited by the coherence length \cite{trovatelloQuasiphasematchedDownconversionPeriodically2025}. 

The direct generation of entanglement in one degree of freedom also opens up new possibilities for designing sources of complex states showing hyperentanglement. For instance, correlations in orbital angular momentum modes with LN have been shown using separable states in the polarization mode\cite{dai_high-dimensional_2022, wu_optical_2023}. Combining this with our approach of directly generating polarization-entangled states may enable the generation of polarization-OAM hyperentangled states.

\begin{acknowledgments}
\noindent This work was supported by the Deutsche Forschungsgemeinschaft (DFG, German Research Foundation) through the International Research Training Group (IRTG) 2675 “Meta-ACTIVE” (project number 437527638), the collaborative research center (SFB) 1375 "NOA"  (398816777), and the MEGAPHONE project (505897284). Also, we acknowledge support by the German Federal Ministry of Research, Technology and Space in the PhoQuant project (FKZ 13N16108) and by the Australian Research Council (\url{https://doi.org/10.13039/501100000923}) Centre of Excellence for Transformative Meta-Optical Systems - TMOS (CE200100010). Sina Saravi acknowledges funding by the Nexus program of the Carl-Zeiss-Stiftung (project MetaNN, project ID P2022-04-018).
\end{acknowledgments}

\appendix

\section{Methods}

The z-cut thin-film lithium niobate of thickness \qty{8}{\micro\meter} on \qty{500}{\micro\meter} SiO$_2$ was obtained from NanoLN. % Add trademark for nanoln?
The sample is illuminated with a continuous-wave pump laser (Thorlabs FPL785P) at \qty{788}{\nano\meter} focused onto the sample using an aspheric lens of focal length \qty{18.4}{\milli\meter} and $\mathrm{NA}=0.15$ with pump beam waist \qty{3.5}{\micro\meter}. The generated photon pairs are collected by an equivalent aspheric lens. 
We ensure there are no stray photons at the telecom range in the pump beam by adding two short-pass filters (Thorlabs FESH850) before the sample, and after the sample, we block the pump beam using long-pass filters (Thorlabs FELH1500 and 2 $\times$ Thorlabs FELH1150).
The collected photon pairs are sent to a 50:50 fiber beam-splitter and then to the superconducting nanowire single photon detectors (SNSPDs, Single Quantum EOS with timing jitter $\leq$ \qty{15}{\pico\second}) and we use the quTAG multi-channel Time-to-Digital converter to record the coincidence histograms.

For SHG measurements, we use the same setup in reverse, pumping with a tunable femtosecond laser (Coherent Chameleon with optical parametric oscillator Angewandte Physik und Elektronik GmbH APE OPO-X) with pulse width \qty{100}{\femto\second}, repetition rate \qty{80}{\mega\hertz}, at a central wavelength \qty{1576}{\nano\meter} and with FWHM \qty{10}{\nano\meter} and the second harmonic is detected using a sCMOS camera (Excelitas pco.edge 4.2 bi). For characterizing the polarization dependence of the sample, we add a half-wave plate (Thorlabs AHWP05M-1600) before the sample and a polarizer (Thorlabs WP25M-UB) before the camera. 

For measuring the entanglement, we perform quantum state tomography on the photon pairs (Fig.~\ref{fgr:2}a). The setup is modified by adding a half-wave plate (Thorlabs AHWP05M-950) in the pump path before the sample to control the linear polarization angle of the pump beam. In the collection arm, we remove the longpass filter Thorlabs FELH1500 and add a dichroic mirror (filter Edmund Optics \#84-656 with cut-off wavelength \qty{1600}{\nano\meter} acting as a dichroic mirror) to spectrally separate the generated biphotons with lower wavelengths in the transmission path and higher wavelengths in the reflection path. In each path, we have a quarter-wave plate (Thorlabs AQWP05M-1600 in the reflection arm and AQWP05M-1430 in the transmission arm), a half-wave plate (Thorlabs AHWP05M-1600) and then a polarizer (Thorlabs WP25M-UB) to select the polarization state to be measured in the arm. Each path is then coupled to a single-mode fiber (SMF-28) and the collected photons are sent to the SNSPDs. We later also add a quarter-wave plate (Thorlabs SAQWP05M-1700) in the pump path, after the pump half-wave plate, to control the ellipticity of the pump beam.

%%%%%%%%%%%%%%%%%%%%%%%%%%%%%%%%%%%%%%%%%%%%%%%%%%%%%%%%%%%%%%%%%%%%%%%%%%%%%%%

\bibliography{bibliography}

\end{document}